\documentclass[prl,twocolumn,floatfix,nobibnotes,superscriptaddress,prb]{revtex4}
\usepackage{amsmath}
\usepackage{graphicx}
\newcommand{\tbf}{\boldsymbol}

\begin{document}
\title{\textit{Ab initio} study of interacting lattice
           vibrations and stabilization of the $\beta$-phase
           in Ni-Ti shape-memory alloy }
\author{Petros Souvatzis}
\affiliation{ Theoretical Division, Los Alamos National Laboratory,
Los Alamos, New Mexico 87545, USA}
\author{Dominik Legut}
\affiliation{Department of Physics and Astronomy, Division of Materials Theory, Uppsala University,
Box 530, SE-75121, Uppsala, Sweden}
\affiliation{Institute of Physics of Materials, Academy of Sciences of the Czech
Republic, v.v.i., \v{Z}i\v{z}kova 22, CZ-616 62 Brno, Czech Republic}
\author{Olle Eriksson}
\affiliation{Department of Physics and Astronomy, Division of Materials Theory Uppsala University,
Box 530, SE-75121, Uppsala, Sweden}
\date{\today }
\author{Mikhail I. Katsnelson}
\affiliation{Institute for Molecules and Materials, Radboud University Nijmegen, NL-6525 ED Nijmegen, The Netherlands}
\begin{abstract}
Lattice dynamical methods used to predict  phase-transformations
in crystals typically evaluate the harmonic phonon spectra and
therefore do not work in frequent and important situations where
the crystal structure is unstable in the harmonic approximation,
such as the $\beta$ structure when it appears as a
high-temperature phase of  the shape memory alloy (SMA) NiTi. Here
it is shown by self consistent {\it ab initio} lattice dynamical
calculations (SCAILD)  that the critical temperature for the
pre-martensitic  $R$ to $\beta$ phase-transformation in NiTi can
be effectively calculated with good accuracy, and that the
$\beta$-phase is a result primarily of the stabilizing  interaction between
different lattice vibrations.
\end{abstract}

\maketitle

\section{I. Introduction}
Shape memory alloys (SMA) are compounds that after a mechanical
deformation can, through heating, retain their original shape
\cite{SMA}. Due to their vast utilization in a broad spectrum of
technologies, spanning areas such as medical applications to
aerospace industry, there is an increased  need for effective
theoretical tools in the development and understanding of these
alloys. Lately several theoretical studies have been made on one of the most commonly  used SMA's, NiTi (nitinol),  focusing  on the 
martensitic transformation path \cite{hatcher1,hatcher2} and on the shape memory behavior \cite{Rabe}.
Here the  theoretical study of NiTi will be continued by applying the recently developed self
consistent {\it ab initio} lattice dynamical method (SCAILD)\cite{petros1}.

The shape memory effect in NiTi is related to a reversible
martensitic phase transformation into a monoclinic structure
(P2$_1$/m, space group 11, Pearson symbol mP4) also known as B19'
phase \cite{SMA2} at around 273 K \cite{SMA3,SMA4,SMA5}. This
phase transformation is preceded by a transformation at about 338
K from the austenite cubic phase (also known as the B2 or
$\beta$-phase, Pm\={3}m, space group 221, Pearson symbol cP2) into
the $R$-phase (P3, space group 143) \cite{ExpNiTi}. The
mechanism behind the $R$ to $\beta$ transformation has been
ascribed  to the suppression of  Fermi surface nesting, resulting in a
hardening of the T$_{2}$A phonon mode at the wave vector (and also
nesting vector) $q=(\frac{1}{3}, \frac{1}{3}, 0)$ \cite{SMEAR}.

Here, by means of first principles calculations, an alternative 
picture of the mechanism  behind this pre-martensitic phase
transition will be provided. We will demonstrate that it is the
interaction between different phonon modes  that provides the main
driving mechanism behind the stabilization of the $\beta$-phase
relative to the $R$-phase in NiTi. Since the $\beta$-phase is
dynamically unstable in the harmonic approximation over a large range of frequencies, not only at the
wave vector $q=(\frac{1}{3}, \frac{1}{3}, 0)$ \cite{PAR,XIA}, it is
absolutely necessary to include anharmonic  effects in any type of
theoretical consideration  when trying to understand the $\beta$
to $R$ phase-transformation in NiTi.

 A straightforward calculation using first-principles molecular dynamics
(MD)~\cite{PARINELLO} should in principle be able to reproduce the
stability of the $\beta$-phase for NiTi, since MD implicitly
include anharmonic effects. However, MD suffers from that it 
is a computationally very demanding task to obtain reliable free energies.
Instead we will make use of the second order
nature of the displacive $\beta$ to $R$ phase-transformation
\cite{ExpNiTi} and take the T$_{2}$A phonon mode displacement at
the wave vector ${\bf q}=[\frac{1}{3} \frac{1}{3} 0]$ as an order
parameter. This will enable us to use the  temperature dependence
of the phonon mode in order to determine  the critical temperature
for phase-transformation.

\section{II. Method}

In order to describe properly the phase-transformation into the
cubic phase for NiTi  one must include the interaction between
phonons~\cite{ktsn}. As a result, phonon frequencies turn out to
be temperature dependent which we explore numerically in this
study by means of the SCAILD method
\cite{petros1,petros2}.

The SCAILD method is based on the calculation of Hellman-Feynman
forces on atoms in a supercell. The  method can be viewed as an
extension of the frozen phonon method \cite{FP1}, in which all
phonons with wave vectors $\mathbf{q}$ commensurate with the
supercell are excited together in the same cell by displacing
atoms situated  at the undistorted positions
$\mathbf{R}+\mathbf{b}_{\sigma}$, according to
 $\mathbf{R}+\mathbf{b}_{\sigma}  \rightarrow \mathbf{R}+\mathbf{b}_{\sigma} + \mathbf{U}_{\mathbf{R}\sigma}$, where the displacements are given by
\begin{equation}
\mathbf{U}_{\mathbf{R}\sigma}
= \frac{1}{\sqrt{N}}\sum_{\mathbf{q},s}\mathcal{A}_{\mathbf{q}s}^{\sigma}
\mathbf{\epsilon}_{\mathbf{q}s}^{\sigma}e^{i\mathbf{q}(\mathbf{R}+\mathbf{b}_{\sigma})}.\label{eq:SUPERPOS}
\end{equation}
Here $\mathbf{R}$ represents the $N$ Bravais lattice sites of the
supercell, $\mathbf{b}_{\sigma}$ the position of atom $\sigma$
relative to this site, $\mathbf{\epsilon}_{\mathbf{q}s}^{\sigma}$
are the phonon eigenvectors corresponding to the phonon mode, $s$,
and the mode amplitude $\mathcal{A}_{\mathbf{q}s}^{\sigma}$  is calculated from the
different phonon frequencies
 $\omega_{\mathbf{q}s}$ through
\begin{equation}
 \mathcal{A}_{\mathbf{q}s}^{\sigma} =\pm \sqrt{
\frac{\hbar}{2M_{\sigma}\omega_{\mathbf{k}s}}coth \Big
(\frac{\hbar\omega_{\mathbf{q}s}}{2k_{B}T} \Big )},
\label{eq:AMPL}
\end{equation}
where $T$ is the temperature of the system.
Here the phonon frequencies
\begin{equation}\label{eq:FOURIER}
\omega_{\mathbf{q}s} =
\Big [ -\sum_{\sigma}\frac{\mathbf{\epsilon}_{\mathbf{q}s}^{\sigma} \cdot \mathbf{F}_{\mathbf{q}}^{\sigma}}{\mathcal{A}_{\mathbf{q}s}^{\sigma}M_{\sigma}}
\Big ]^{1/2},
\end{equation}
are obtained from the Fourier transform
$\mathbf{F}_{\mathbf{k}}^{\sigma}$ of the forces acting on the
atoms in the supercell.

Due to the simultaneous presence of all the commensurate phonons
in the same force calculation, the interaction between different
lattice vibrations is  taken into account and the phonon
frequencies given by  Eqn. \ref{eq:FOURIER} are thus renormalized
by the very same interaction.

By alternating between calculating the forces on the displaced
atoms and calculating new phonon frequencies and new displacements
through Eqn.\ref{eq:SUPERPOS}- \ref{eq:FOURIER} the phonon
frequencies are calculated in a self consistent manner. For more
details on the SCAILD method we refer to Refs.
\cite{petros1,petros2,petros3}.

It should be mentioned that we do not consider here the phonon
decay processes (see, e.g., Ref.~\cite{CaSr} and Refs. therein).
Thus the  question of  how  phonon line widths obtained within the  SCAILD framework 
are related to experimentally observed  line widths is still an open question.
In the present  calculations thermal expansion effects have not
been taken into account, and all calculations have been performed
at the constant experimental lattice constant of 3.01 {\AA}
\cite{SMA3, lattc}. 
 
 As regards the computational details of the force calculation we
used the VASP package \cite{VASP}, within the generalized gradient
approximation (GGA). The PAW potentials used required energy
cutoffs of 300 eV. Methfessel-Paxton smearing of 0.2 eV was used together with a $8\times8\times8$ Monkhorst-Pack k-point
grid. The supercell used was obtained by increasing the cubic
primitive cell 3 times along the 3 primitive lattice vectors,
resulting in a 54 atom supercell. Furthermore, the frozen-phonon
calculations were performed with $1\times3\times3$ supercells utilizing
$45\times15\times15$ Monkhorst-Pack k-point grids, whereas the Fermi surfaces
and general susceptibilities were calculated using a $100\times100\times100$
Monkhorst-Pack mesh.

\begin{figure}[tbp]
\begin{center}
\includegraphics*[angle=0,scale=0.49]{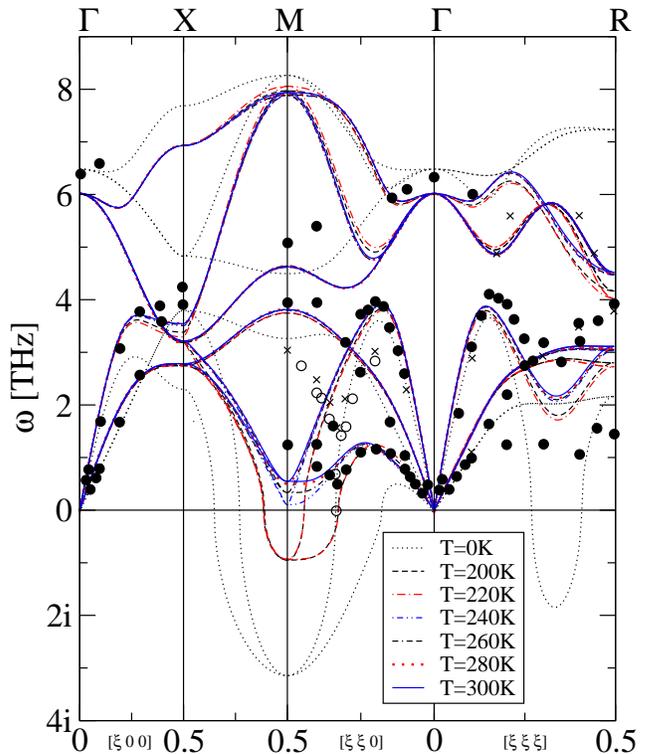}
\caption{(Color online) The phonon dispersions of $\beta$-NiTi calculated at different temperatures together with experimental  data  measured at 400 K (black circles), at 338 K (empty circles) and at 423 K (crosses)
\cite{ExpNiTi}. Solid, dashed, dotted and dashed-dotted lines are the first principles self consistent 
phonon calculations.}
\label{fig:firstp}
\end{center}
\end{figure}
\begin{figure}[tbp]
\begin{center}
\includegraphics*[angle=0,scale=0.32]{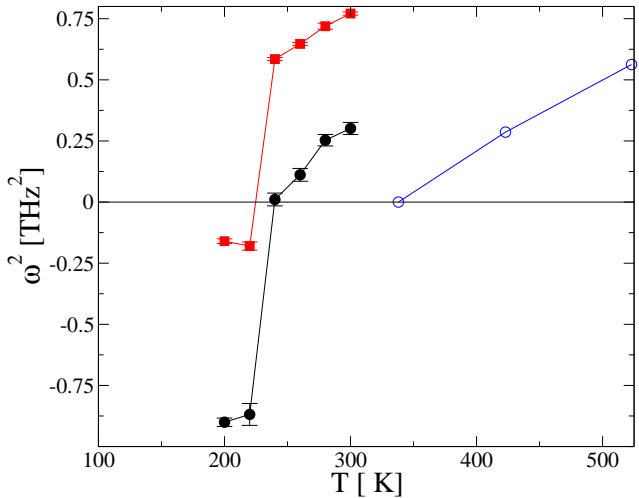}
\caption{(Color online) The calculated temperature dependence of the T$_{2}$A phonon frequency  at $q=[\frac{1}{3} \frac{1}{3} 0]$ (red squares) and at  $q=[\frac{1}{2} \frac{1}{2} 0]$ (black circles) in  $\beta$-NiTi, here displayed together with experimental data for $q=(\frac{1}{3},\frac{1}{3},0)$ (empty blue circles) \cite{ExpNiTi}. The width of the error bars are the square root of the mean square deviation of the last 10 SCAILD-iterations relative to the frequency of the 150th SCAILD-iteration.}
\label{fig:temp}
\end{center}
\end{figure}
\begin{figure}[tbp]
\begin{center}
\includegraphics*[angle=0,scale=0.2075]{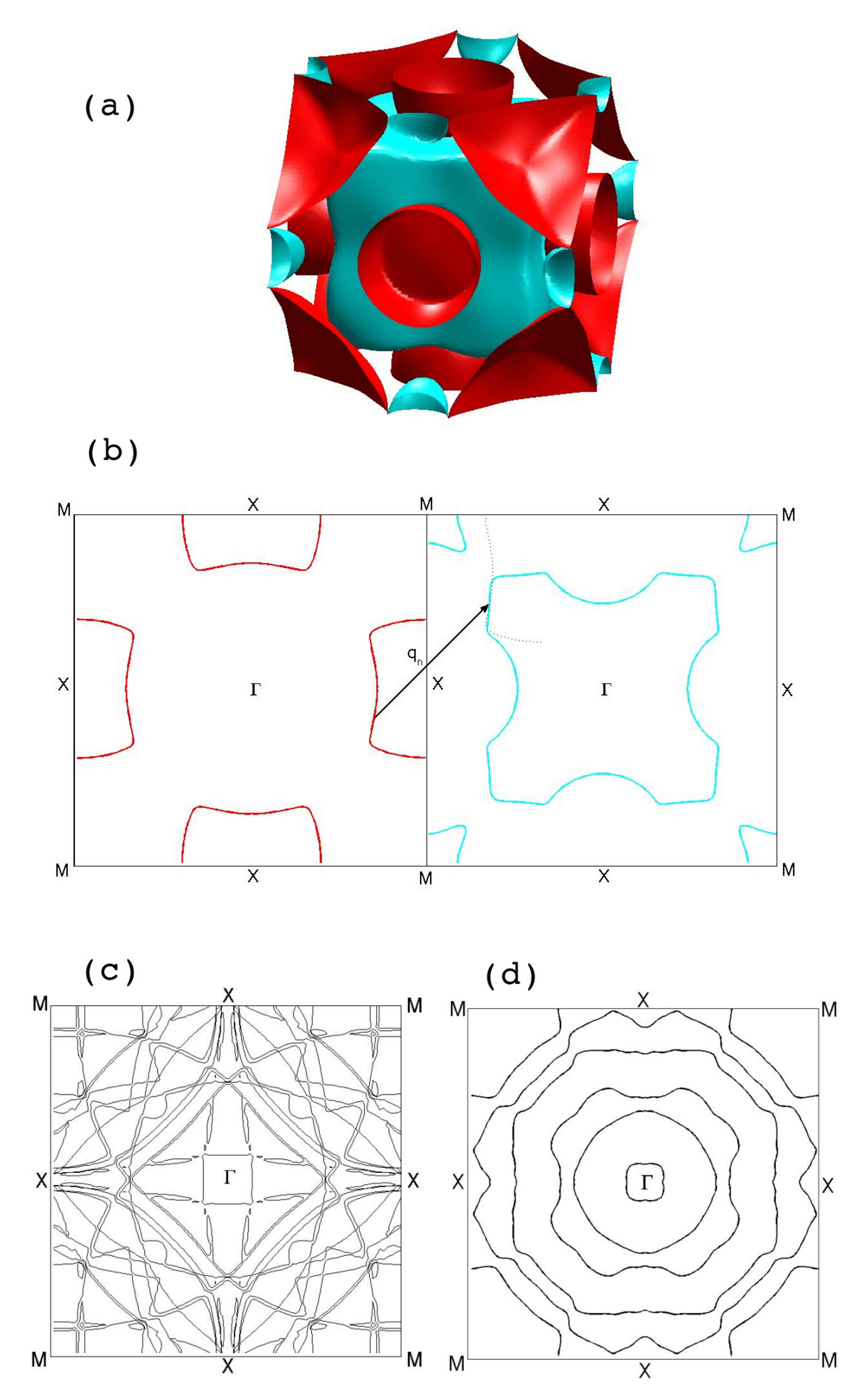}
\caption{(Color online) Fermi surface of the $\beta$-NiTi. In (a) the Fermi surface in the T=0 case  (i.e no phonon induced disorder). In (b) we show the cut through the Fermi surface displayed in (a). 
The left most panel in (b) shows a cut through the red bowl shaped  surface sheets  in (a). The right most panel in (b) shows a cut through the turquoise sheets in (a). In (b) the nesting vector $q_{n}=(\frac{1}{3},\frac{1}{3},0)$ interconnecting the nested parts of the Fermi surface is also shown. In (c) we show
a cut through of the Fermi surface in (a) down-folded to the 1st Brillouin-zone of the undistorted $3\times 3\times 3$ super cell. In (d) we show a cut through the Fermi surface calculated from four of the T=300K atomic configurations produced by the SCAILD scheme.}
\label{fig:FS}
\end{center}
\end{figure}
\begin{figure}[tbp]
\begin{center}
\includegraphics*[angle=0,scale=0.3]{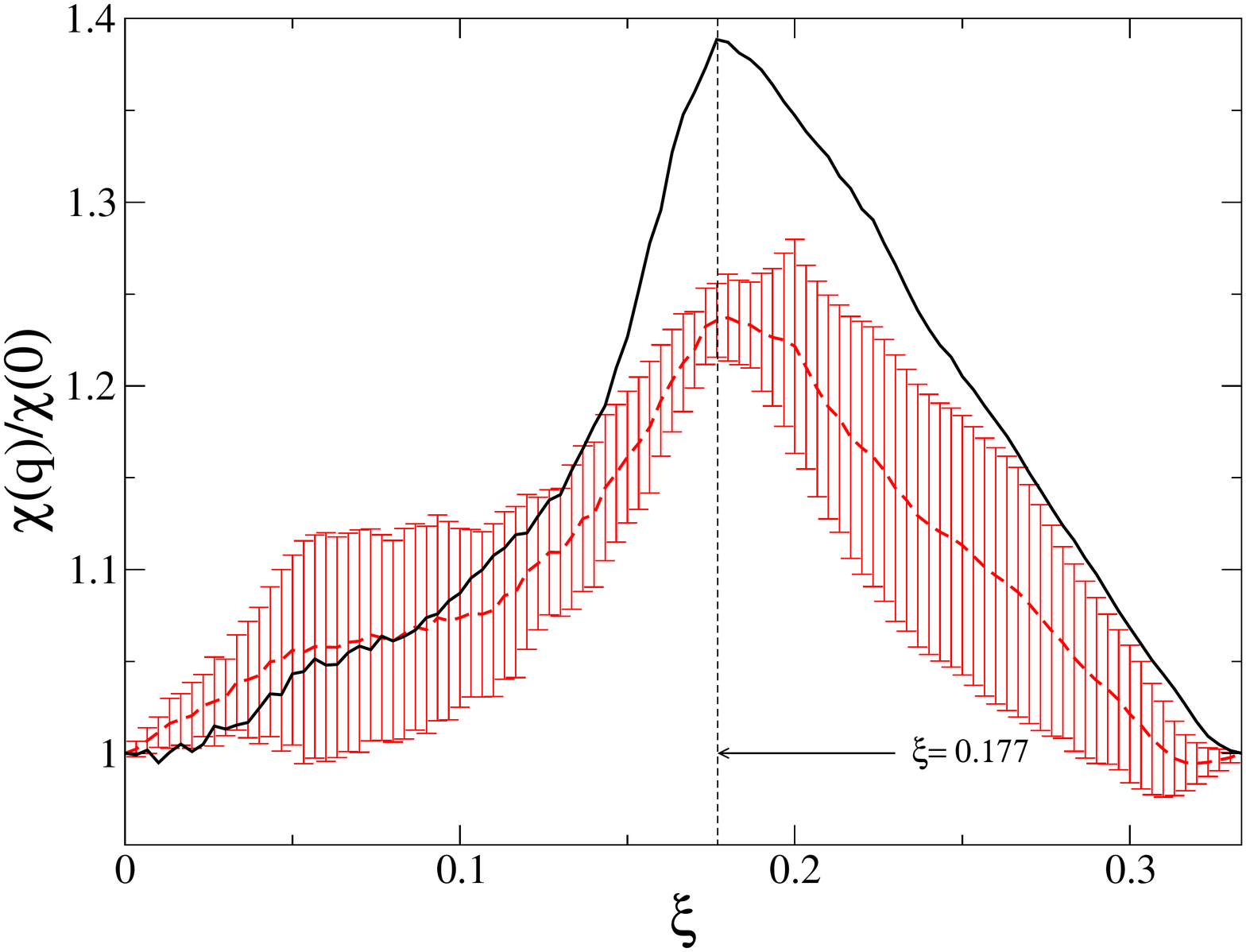}
\caption{(Color online) The calculated susceptibility as a function of $q=(\xi,\xi,0)$ in
 $\beta$-NiTi. The full black curve is  the susceptibility for the T= 0 K case (i.e no phonon induced disorder). The dashed red curve is  the mean susceptibility calculated from four of the T=300K atomic configurations produced by the SCAILD scheme. Here the width of the error bars correspond to the standard  deviation of the T=300K susceptibility distribution.
 The susceptibilities are calculated within a $3\times3 \times 3$ supercell, resulting in a shift  of the susceptibility peaks from $\xi=\frac{1}{3}$ to $\xi\approx0.177$, due to the down-folding of the bands into the 1st Brillouin zone of a $3\times3 \times 3$ cell.}
\label{fig:SUS}
\end{center}
\end{figure}
\section{III. Results}
Fig. \ref{fig:firstp} shows the calculated phonon dispersions in
cubic NiTi  for the temperatures 0 K, 200 K, 220 K, 240 K, 260 K,
280 K and 300 K. The phonon dispersion relation at T=0K is very
similar to the previous calculation done in Ref.\cite{PAR},
including imaginary frequencies along both directions (i.e.
[$\xi$,$\xi$,0] and [$\xi$,$\xi$,$\xi$]). The finite temperature
calculations predict the stability of the cubic  phase of  NiTi by
promoting the frequencies of the phonons along the $\Gamma$ to $R$
symmetry line and around the $M$ symmetry point from imaginary to
real for temperatures $\gtrsim$ 238 K.

Furthermore, the calculated T=300K phonon dispersion is
in good agreement with the experimental T=400K data (black circles), with the exception of the 
lowest lying acoustic branch along the $\Gamma$ to $R$ symmetry line.

Fig. \ref{fig:temp} shows the  calculated squared T$_{2}$A phonon frequency at the wave vectors  $q=(\frac{1}{2},\frac{1}{2}, 0)$ and $(\frac{1}{3},\frac{1}{3},0)$ at different temperatures together with experimental data. Here, as a result of  a fourth order anharmonic interaction, the expected linear dependence $\omega^{2}\sim T$  \cite{Lin}, also suggested by experiment, is reproduced by the calculation.

 The sudden jump in the calculated squared T$_{2}$A phonon frequencies  at  T$\sim 227$ K (Fig. \ref{fig:temp}) can be related to the limited size of the 
supercell, since it overestimates the different phonon mode contributions to the atomic displacements  $\sim \frac{1}{\sqrt{N}\omega}$,  especially  in the temperature range where $\omega$ is close to zero.
Thus by increasing the size of the super cell, i.e increasing the number of commensurate phonons, this overestimation can at least in principle be avoided. Furthermore, in the calculated phonon dispersion (Fig. \ref{fig:firstp}) the dip or singularity in the T$_{2}$A phonon frequency  is shifted from the experimental position $q=(\frac{1}{3},\frac{1}{3},0)$ to $q=(\frac{1}{2},\frac{1}{2}, 0)$. This shift  originates from the singularity being  confined to a relatively small region of q-space which also  cannot be described adequately by a small supercell \cite{KATS}.
However, increasing the currently 
used $3\times 3 \times 3$ supercell to the smallest  larger cell accommodating the  q=$(\frac{1}{3}, \frac{1}{3}, 0)$ wave vector, would imply the use of a 6x6x6 cell  which was not pursued, due to computational reasons.

By using the phonon frequency of the T$_{2}$A mode at $q=(\frac{1}{3}, \frac{1}{3},0)$ as an order parameter for the $\beta$ to $R$ phase-transformation the critical temperature, $T_{c}$,
corresponding to the transformation can be estimated to $\sim$ 227 K. However, if  instead  the T$_{2}$A phonon frequency at  $q=(\frac{1}{2},\frac{1}{2}, 0)$ is used as a order parameter  we have $T_{c}\sim$ 238 K.  This should be compared to the experimental value of 338 K. Since $T_{c}$ depend
 strongly on the alloy composition (a change from 50 at.\% to 51 at.\% Ni lowers $T_{c}$ with up to 100 K) \cite{dependence1} and on oxygen and carbon impurities \cite{dependence2}, the agreement must be viewed as good. 

To investigate the relative importance between two of the possible
processes involved in stabilizing the T$_{2}$A mode at  $q=(\frac{1}{3}, \frac{1}{3},0)$: (1) destruction of  Fermi surface nesting through the 
thermal smearing related to  electronic excitations (used by Zhao {\it et al} \cite{SMEAR} to illustrate the effect of nesting suppression), or, (2) phonon-phonon
interactions, a series of  additional first principles electronic structure
calculations were performed.  First,  the frequency of the
T$_{2}$A-mode was calculated at different thermal smearings of the
electronic subsystem,  through a series of frozen phonon
calculations \cite{FP1}. The results of these calculations revealed 
that temperatures  above 1000 K,  i.e. much higher than the observed transition temperature,  were required if  thermal smearing
was to be the only effect responsible for the  stabilization of the
the T$_{2}$A  phonon mode.

In the second step a series of Fermi surface calculations were performed for the $\beta$-NiTi  phase for different phonon excited geometries. In Fig. \ref{fig:FS}(a)  the Fermi surface of  $\beta$-NiTi  with  no phonon induced 
atomic disorder is shown. In Fig \ref{fig:FS}(b) a cut,  in the plane $k_{z}=0$, through the  same Fermi surface as  in 
Fig. \ref{fig:FS}(a) is shown, illustrating the nesting features.  
The Fermi Surface of the undistorted $\beta$-structure was
also calculated within a  $3\times 3 \times 3$ supercell, the result is displayed in Fig. \ref{fig:FS}(c).
In Fig. \ref{fig:FS}(d) a cut through the Fermi surface is shown that has been calculated from four of the atomic configurations produced by the SCAILD scheme at T=300 K. 
This cut was taken through the surface $\langle E(\mathbf{k})\rangle = \langle E_{F}\rangle$, where  $\langle E(\mathbf{k})\rangle$ and 
$\langle E_{F}\rangle$ are the arithmetical mean values of the Kohn-Sham eigenvalues, $E(\tbf{k})$, and Fermi levels, $E_{F}$, calculated from 
the atomic configurations produced by the SCAILD scheme at T=300 K. Here $\mathbf{k}$ denotes a point in the space of k-points.

 Due to
the down-folding of the bands in Fig. \ref{fig:FS}(c-d), the nesting vector $q_{n} =(\frac{1}{3},\frac{1}{3},0)$ is shifted to $q=(0.177,0.177,0)$. Fig. \ref{fig:FS}(c-d) shows an apparent change of  the Fermi surface topology as  the phonon induced atomic disorder is introduced. However, to properly gauge the
effect of atomic disorder upon the 
nesting features of $\beta$-NiTi,  the susceptibility \cite{KATS}
 \begin{equation}\label{eq:susept}
\chi(\mathbf{q})=\sum_{\mathbf{k}}\sum_{n,m}\frac{f[E_{n}(\mathbf{k+q})]-f[E_{m}(\mathbf{k})]}{E_{n}(\mathbf{k+q})-E_{m}(\mathbf{k})}
 \end{equation}
 was also calculated for the same atomic configurations as was used in the Fermi surface calculations. Here 
 $f(E)$ is the Fermi-Dirac distribution function given by $f(E) = 1/(e^{\frac{E-E_{F}}{k_{B}T}}+1)$. In Fig. \ref{fig:SUS} the results of these calculations are displayed, showing the suppression of the susceptibility peak as the phonon induced atomic disorder is introduced.  It should be noted that the suppression of the peak in x(q) is quite pronounced. This demonstrates that the basic electronic structure of NiTi is drastically different when the finite temperature excites collective lattice vibrations, compared to a T=0 calculation. Hence, it is this change in the electronic structure and the accompanying modification of the force constant matrix which is primarily responsible for the stabilization of the beta-phase. This explanation is hence somewhat more intricate and complex than the conventional model, of a smearing of a rigid electronic structure due to temperature effects of the Fermi-Dirac distribution function.
 
\section{IV. Conclusion}
To summarize, by first principles SCAILD calculations, the cubic $\beta$ phase
in NiTi has been shown to be stabilized by phonon-phonon
interactions. Also, in the case of the unstable T$_{2}$A phonon mode at q=$(\frac{1}{3},\frac{1}{3},0)$ this interaction has been shown to be mediated through thermal disorder induced suppression of  Fermi surface nesting. 

Furthermore, the SCAILD method has been proven  an accurate and
effective theoretical tool by predicting the critical temperature
between 227 K $\lesssim T_{c} \lesssim$ 238 K for the $\beta$ to
$R$ pre-martensitic phase-transformation, which is comparable with
the experimental value of $T_{c}\sim$ 338 K. \cite{ExpNiTi} 

The Department of Energy supported this work under Contract
No.~DE-AC52-06NA25396.
We thank Bruce N. Harmon for all the helpful discussions.

\end{document}